\documentclass[a4paper,12pt]{article}
\usepackage{amsmath,amssymb,dsfont}
  \usepackage{paralist}
  \usepackage{graphics} 
  \usepackage{epsfig}
 \usepackage[colorlinks=true]{hyperref}

\newcommand{\diff}{\psi_{\pm}}

\newcommand{\R}{\mathbb{R}}

\title{The 2-component dispersionless Burgers equation arising in the modelling of blood flow}

\author{Tony Lyons}
\date{}

\begin{document}

\maketitle

\centerline{Tony Lyons}
\medskip
{\footnotesize
 \centerline{School of Mathematical Sciences, Dublin Institute of Technology}
  \centerline{Kevin Street, Dublin 8, Ireland}
} 



\begin{abstract}
 This article investigates the properties of the solutions of the dispersionless
two-component Burgers (B2) equation, derived as a model for blood-flow in arteries with elastic walls. The phenomenon of wave breaking is investigated as well as applications of the model to clinical conditions.\footnote{The author is grateful to Prof. A. Constantin for many helpful suggestions during his visit to ESI-Vienna. This material is based upon works supported by the Science Foundation Ireland (SFI), under Grant No. 09/RFP/MTH2144.}

\end{abstract}

\section{Introduction}
There are many examples of systems of nonlinear PDEs modelling the propagation of waves in fluids under various conditions. One of the earliest models was of course the KdV equation, modelling the propagation of surface waves in shallow water. Meanwhile the nonlinear Schr\"{o}dinger equation has proved very successful at modelling the propagation of wave fronts in deep water.

The Camassa-Holm equation \cite{CH93}
\begin{equation}\label{eq1}
 u_{t}-u_{xxt}+2\omega u_{x}+3uu_{x}-2u_{x}u_{xx}-uu_{xxx}=0,
\end{equation}
has gained popularity as an integrable model with many applications and interesting
properties \cite{Beals_et_al, CGI06, BKST09}. Among its many applications, it is well known as a model of shallow water waves, admitting many novel solutions cf. \cite{Johnson02, CL09}. Another physical application of the system arises in the study of axially symmetric deformation waves propagating in hyperelastic rods cf. \cite{Dai98, CS00}.
The system also admits various generalisations, the most popular of which is the two-component
one (CH2) \cite{SA}:
\begin{eqnarray}
m_{t}+2mu_{x}+um_{x}+\rho\rho_{x}=0,\label{ch2_1}\\
\rho_{t}+\left(\rho u\right)_{x}=0.\label{ch2_2}
\end{eqnarray}

\noindent where $m=u-u_{xx}$. Taking $\rho=0$ reduces the CH2 system to the CH equation
(\ref{eq1}). The CH2 energy Hamiltonian is given by
\begin{equation}\label{im}
H_1=\frac{1}{2}\int(um+\rho^2)\text{d}x,
\end{equation}
and is clearly positive-definite. The CH2 system (\ref{ch2_1})-(\ref{ch2_2}) is  bi-Hamiltonian. This means it possesses two compatible Poisson brackets. The first Poisson bracket between two functionals $F$ and $G$ of the variables $m$ and $\rho$ is in semidirect-product Lie-Poisson form \cite{SA, CI08,I09}:
\begin{equation}\label{ipb}
\{F,G\}_1=-\int\bigg[\frac{\delta F}{\delta m}(m\partial+\partial
m)\frac{\delta G}{\delta m}
+\frac{\delta F}{\delta m}\rho\,\partial\frac{\delta G}{\delta \rho}
+\frac{\delta F}{\delta \rho}\partial \rho \frac{\delta G}{\delta m}
\bigg]\text{d}x.
\end{equation}
This Poisson bracket generates the CH2 system from the Hamiltonian
$H_1$. Its second Poisson
bracket has constant coefficients,
\begin{equation} \{F,G\}_2=-\int\Big[\frac{\delta F}{\delta
m}(\partial-\partial^{3})\frac{\delta G}{\delta m}+\frac{\delta
F}{\delta \rho}
\partial\frac{\delta G}{\delta \rho}\Big]\text{d}x,  \label{pb2}
\end{equation}
and corresponds to the Hamiltonian $H_2=\frac{1}{2}\int
(u\rho^2+u^3+uu_x^2)\text{d}x$. There are two Casimirs for the second bracket: $\int \rho\, \text{d}x$ and $\int m\, \text{d}x$.

The CH2 system was initially introduced in \cite{SA} as a tri-Hamiltonian (integrable) system, and was studied further by others, see, e.g.,
\cite{Escher07,H09,LZ05,CLZ05,F06,I06,CI08,HT09,GL10}. The inverse scattering transform for the CH2 system is developed in \cite{HI11}.

The CH2 model has various applications. For example, in the context of shallow water waves propagating over a flat bottom, $u$ can be interpreted as the horizontal fluid velocity and $\rho$ is the water elevation in the first approximation \cite{CI08,I09}. In the shallow water regime there are scale
characteristics, one of which is $\delta=h/\lambda$ where $h$ is the average
depth and $\lambda$ is the wavelength of the water waves.
With these scale factors, if $u=\mathcal{O}(1),$ then $m=u-\delta^2 u_{xx}$,
see e.g. \cite{CI08,I09}. In the limit $\lambda>> h$ or $\delta\to 0,$ we
have $m=u$, which we call the `two component dispersionless
Burger's equation':
\begin{eqnarray}
u_{t}+3uu_{x}+\rho\rho_{x}=0,\label{b2_1}
\\
\rho_{t}+\left(\rho u\right)_{x}=0.\label{b2_2}
\end{eqnarray}

The first Poisson bracket (\ref{ipb}) remains valid in the case $m=u$ with Hamiltonian $H_1$.
The second Poisson bracket (\ref{pb2}) in this limit
becomes
\begin{equation}
\{F,G\}_2=-\int\Big[\frac{\delta F}{\delta
u}\partial\frac{\delta G}{\delta u}+\frac{\delta
F}{\delta \rho}
\partial\frac{\delta G}{\delta \rho}\Big]\text{d}x,  \label{pb2B2}
\end{equation}
and
the corresponding Hamiltonian is $H_2=\frac{1}{2}\int
(u\rho^2+u^3)\text{d}x$. Dispersionless systems like (\ref{b2_1})-(\ref{b2_2}) also arise in the context of very long water waves, in particular in the modelling of tsunamis as they approach the shore \cite{CJ2008}.

\begin{figure}[h]
\begin{center}
\includegraphics*[width=0.875\textwidth]{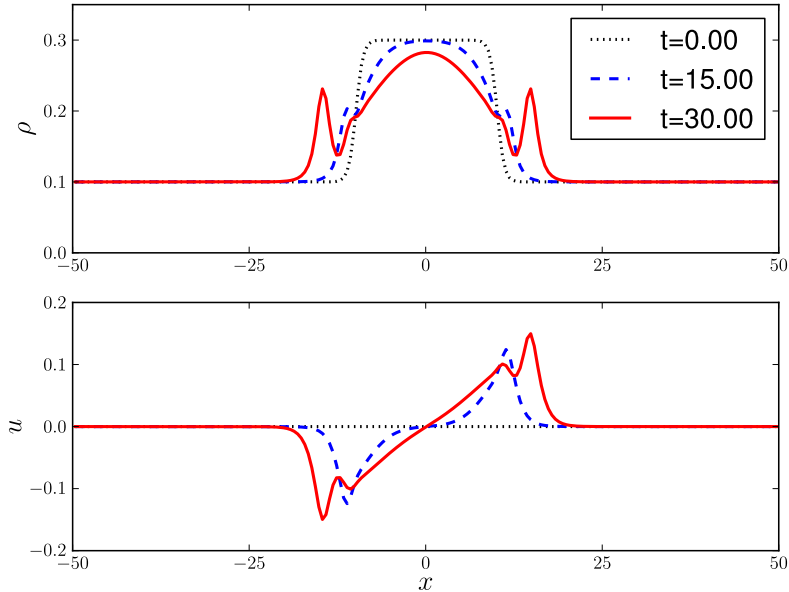}
\end{center}
\caption{\label{CH2-figs} Dam-break results for the CH2 system in equations (\ref{ch2_1})-(\ref{ch2_2}) show evolution of the elevation $\rho$ (upper panel) and velocity $u$ (lower panel), arising from initial conditions
(\ref{DB1})- (\ref{DB2}) in a periodic domain. The soliton solutions are seen to emerge after a finite time, and the evolution of both variables generates more and more solitons propagating in both directions as time progresses.   Figures are courtesy of J. Percival. }
\end{figure}

\begin{figure}[h]
\begin{center}
\includegraphics*[width=0.875\textwidth]{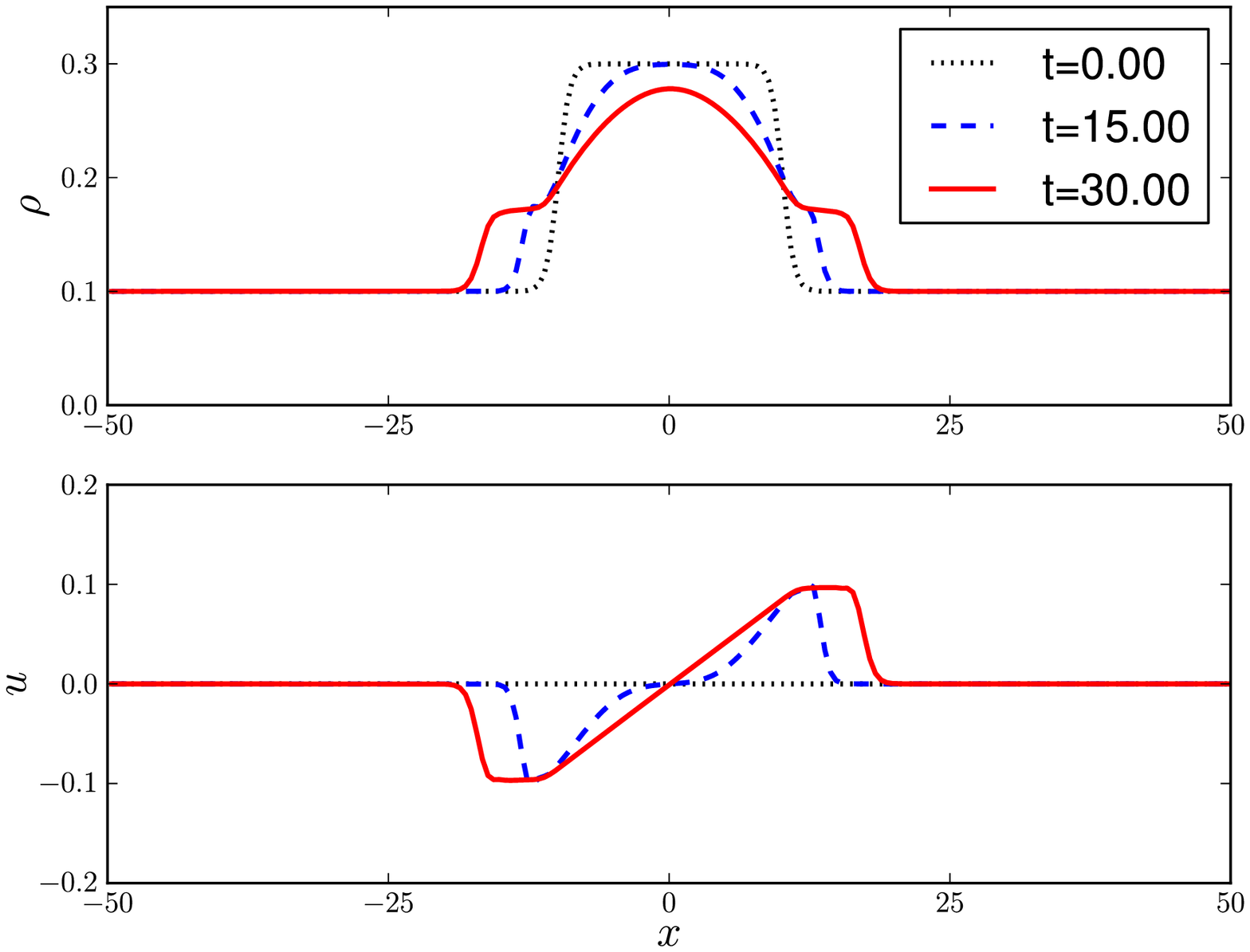}
\end{center}
\caption{\label{B2-figs} Dam-break results for the system (\ref{b2_1})-(\ref{b2_2}) shows evolution of the elevation $\rho$ (upper panel) and velocity $u$ (lower panel), arising from initial conditions (\ref{DB1})- (\ref{DB2}) in a periodic domain. 
observe a wave with a slope that is becoming closer to vertical which eventually
 forms a breaking wave.  Figures are courtesy of J. Percival. }
\end{figure}

Both systems are illustrated in Fig.\ref{CH2-figs} and Fig.\ref{B2-figs}
with the so-called `dam-break' initial conditions. The dam-break problem involves a body of water of uniform depth, initially retained behind a barrier.  When the barrier is suddenly removed at $t=0$, the water flows downward and outward under gravity.  The problem is to find the subsequent flow and determine the shape of the free surface.  This question is addressed in the context of shallow-water theory, e.g., by Acheson \cite{Ach1990}, and thus serves as a typical hydrodynamic problem of relevance for shallow
water equations. Dam break initial conditions are:
\begin{eqnarray}
u\left(x,0\right)&=&0,\label{DB1} \\
\rho\left(x,0\right)&=&0.1\left(1+\tanh\left(x+5\right)-\tanh\left(x-5\right)\right),\label{DB2}
\end{eqnarray}
solved in a periodic domain of width 100 for the CH2 and Burger's
equations.

Both are solved in conservative form using an implicit midpoint rule.
For CH2 the momentum equation is written as
\[\left(\frac{m}{\rho}\right)_{t}+\left(\frac{mu}{\rho}\right)_{x}+\rho_{x}=0.\]
For Burger's equation, a potential form is used
\[\left(\rho\phi\right)_{t}+\frac{3}{2}\left(\rho u\phi\right)_{x}+\frac{\rho^{3}}{2}=0,\]
\[u=\phi_{x}.\]

In Section 2 we develop a model of blood flow in elastic tubes modelling arteries. The blood is modelled as a Newtonian fluid, which is a good approximation in the case of the main arteries, as the diameter of the artery is much larger than the cross-section of a typical blood cell. In Section 3 we show that the model developed in Section 2 may be solved directly by using characteristic curves. Section 4 examines a related model, however unlike the primary model we work with, the model examined in Section 4 does not possess a positive definite Hamiltonian. We find explicit solutions to this system based on a method analogous to the method o characteristic outlined in Section 3. In Section 5 we show that the two component model developed in Section 2 does display the phenomenon of wave breaking, and we give a physiological interpretation of the result based on clinical observations. In this section we also discuss the initial conditions for which the system possess global solutions.

\section{Modeling Newtonian fluids in elastic tubes}
The 2-component Burgers equation, and related systems, are well understood quasi-linear systems arising in many physical applications, for example as model of shock-waves in gas dynamics (see e.g. \cite{Strauss1992}, Section 13.2). In this paper we give a derivation of the system (\ref{b2_1})-(\ref{b2_2}) in the context of a Newtonian fluid within an elastic tube, modelling the flow of blood within arteries.

The model is quasi-linear in the dynamical variables, $u(x,t)$-the fluid velocity, and $A(x,t)$-the tubes cross section.
The physiological relevance of the current article is in relation to the so called \textit{pistol-shot pulse.} This is a popular term for a phenomenon reported by clinicians in which a loud cracking sound is heard by the stethoscope over an artery, caused by a large distension followed by an abrupt collapse of the artery wall. This phenomenon is known to occur in large arteries during aortic regurgitation, a condition in which some blood leaks back into the left ventricle during systole.

To establish a model of arterial blood flow, we consider a simplified model of blood itself, in that it will be modelled as an incompressible, inviscid and irrotational Newtonian fluid. The artery is modeled as an axially symmetric elastic tube, with cross section $A(x,t)$ which depends on the axial coordinate $x$ and time $t.$ Being elastic, the artery experiences a restoring force when expanded from it's equilibrium cross-section, and as such exerts additional pressure on the blood contained within.

The volume of blood contained in the infinitesimal volume of artery between the axial locations $x$ and $x + dx$, at a fixed instant $t$ is $dV = A(x,t)dx.$ In addition, the mass content in that same volume of blood will be the material density at the location at that instant $\tilde{\rho}(x,t)$, times the volume itself, or
\[dM_{(t)} = \tilde{\rho}(x,t)dV  = \tilde{\rho}(x,t)A(x,t)dx.\]
The sub-script $(t)$ on $M_{(t)}$ denotes the mass of the blood at a fixed instant $t$. Since we model the blood as an incompressible fluid, the density of the blood is homogenous throughout the artery at all instants, and so in our model the blood density is a fixed constant $\tilde{\rho}.$ Within a finite volume of artery, between the axial locations $a$ and $x$, with $a < x,$ and at some fixed moment $t,$ it follows that the mass content is,
\[\int dM_{(t)} = \tilde{\rho}\int_{V_{ax}}dV = \tilde{\rho}\int_{a}^{x}A(\xi,t)d\xi.\]
The sub-script appearing in $V_{ax}$ denotes the volume between the axial locations $a$ and $x.$

The velocity of the blood through the artery at a fixed axial coordinate and time will be uniform throughout the cross-section at that location and time. In an infinitesimal time element $dt,$ the blood content at the location $x$ will be displaced by a distance $dX = u(x,t)dt.$ In this time interval, the total mass of blood to traverse the cross-section $A(x,t)$ will be
\[dM = \tilde{\rho}A(x,t)dX = \tilde{\rho}A(x,t)u(x,t)dt.\]
It follows that the change in the blood content in the arterial volume between $a$ and $x,$ in the time interval $dt,$ is the total blood displacement into the volume across $A(a,t)$ less the total blood displacement out of the volume across $A(x,t).$ So we may write,
\[\int_{a}^{x}dM = \tilde{\rho}u(a,t)A(a,t)dt - \tilde{\rho}u(x,t)A(a,t)dt.\]
No subscript appears on the mass element $dM$ on the left hand side above, since we are evaluating the change in mass content over the infinitesimal time interval $dt.$

We may rewrite the above equation as one involving the rate of change of blood content within a finite arterial volume as follows,
    \begin{equation*}\label{mass_cons.1}
        \partial_{t}\left(\int_{a}^{x}dM_{(t)}\right) = \partial_{t}\left(\tilde{\rho}\int_{0}^{x}A(\xi,t)d\xi\right) = \tilde{\rho}u(0,t)A(0,t) - \tilde{\rho}u(x,t)A(x,t).
    \end{equation*}
Operating on this equation with $\partial_{x},$ the fundamental theorem of calculus applied to the left-hand side gives us
\[\partial_{x}\partial_{t}\left(\tilde{\rho}\int_{a}^{x} A(\xi,t)d\xi\right) = \tilde{\rho}\partial_{t}\left(\partial_{x}\int_{a}^{x}A(\xi,t)d\xi\right) = \tilde{\rho}\partial_{t}A(x,t),\]
while $\partial_{x}$ applied to the right hand side gives us
\[\partial_{x}(u(a,t)A(a,t) - u(x,t)A(x,t)) = -\partial_{x}(u(x,t)A(x,t)).\]
Thus we arrive at the continuity equation,
    \begin{equation}\label{gov.1}
        A_{t}(x,t) + (u(x,t)A(x,t))_{x} = 0,
    \end{equation}
where sub-scripts denote partial differentiation with respect to the particular variable. The factor of $\tilde{\rho},$ common among all three terms has been cancelled from the above equation.

The physical interpretation of this equation is as follows. In an infinitesimal time interval $dt,$ the infinitesimal increase (decrease) in the blood content $dM$ at  $x$ causes an increase (decrease) of the blood volume, since the blood itself is incompressible. As the walls of the artery are elastic, this increase (decrease) in blood volume is accommodated by an expansion (contraction) of the artery, which is realised as an increase (decrease) in its cross section $A(x,t),$ at the location $x.$

Newton's laws dictate that any forces acting upon the blood will cause a corresponding change of momentum of the blood. In the case of fluids, the forces acting upon any individual element may be separated into two categories
\begin{itemize}
  \item The external forces acting on the fluid elements, which in this case will be the restoring forces present in the artery walls, thereby exerting an external pressure on the blood at a given location $x$ and some instant $t.$
  \item The internal forces acting upon a fluid element caused by contiguous fluid elements.
\end{itemize}
Both external and internal forces are found to cause an overall change in the linear momentum of an individual fluid element, in accordance with Newton's second law.

The linear momentum of an infinitesimal volume of blood at some location $x$ and instant $t$ is $\tilde{\rho}u(x,t)dV_{x} = \tilde{\rho}u(x,t)A(x,t)dx,$ or the infinitesimal mass element times the velocity thereof. The time rate of change of the momentum, namely $\tilde{\rho}\partial_{t}(u(x,t)A(x,t))dx,$ will appear as the left hand side of Newton's second law. In the case of a finite volume of blood this becomes
\[\tilde{\rho}\partial_{t}\left(\int_{a}^{x}u(\xi,t)A(\xi,t)d\xi\right),\]
which is the rate of change in momentum of the blood contained within the artery between $a$ and $x$ at the instant $t.$
As we have already seen, an infinitesimal time displacement causes a corresponding spatial displacement of the blood at $x,$ by an amount $dX.$ However, in moving this distance the velocity of that blood will change by the amount
\[u_{x}(x,t)dX = u(x,t)u_{x}(x,t)dt.\]
It follows that the corresponding change in linear momentum of this blood will be it mass times the change in velocity, or
\[\tilde{\rho}u(x,t)u_{x}(x,t)A(x,t)dx,\]
which for a finite volume becomes
\[\tilde{\rho}\int_{a}^{x}u(\xi,t)\partial_{\xi}u(\xi,t)A(\xi,t)d\xi.\]
Thus we have established the change in linear momentum of a finite volume of blood, due to the spatial displacement thereof, during an infinitesimal time increment $dt.$

Next we must also include the possibility that the linear momentum of the blood will change in an infinitesimal time $dt,$ due to a change in the blood-content. Suppose at some instant $t$ the rate of increase of blood at $x$ is
\[u(x,t)A(x,t),\]
while the rate of decrease of blood at $x + dx$ is
\[-u(x+dx,t)A(x+dx,t) \simeq -u(x,t)A(x,t) - u_{x}(x,t)A(x,t)dx - u(x,t)A_{x}(x,t)dx.\]
It follows that the rate of change of the mass will be
\[-\tilde{\rho}(u(x,t)A(x,t))_{x}dx,\]
and the corresponding rate of change of the linear momentum will be
\[-\tilde{\rho}(u(x,t)A(x,t))_{x}u(x,t)dx.\]
In the case of a finite volume, the corresponding change in linear momentum is,
\[-\tilde{\rho}\int_{a}^{x}u(\xi,t)(u(\xi,t)A(\xi,t))_{\xi}d\xi.\]
This concludes the effect of internal forces acting upon an individual fluid element within the artery.

Next we need to investigate the effects of external forces on a fluid element. In this case the external forces are provided by the restoring forces within the arterial wall itself, which exerts a force on the blood. Between the axial locations $x$ and $x + dx,$ the internal surface of the artery has a directed area element $\hat{n}dS,$ with $\hat{n}$ being the outward normal and $dS$ being the magnitude of the infinitesimal area element. The force exerted by the tube on the blood is  $-P(x,t)\hat{n}dS,$ where $P(x,t)$ denotes the pressure exerted due to the restoring force of the artery acting on the blood. In the case of a finite volume $V$ between $a$ and $x,$ we find the corresponding force exerted by the artery with surface area $S$ to be
\[-\int_{S} P(\xi,t)\hat{n}dS.\]
Applying Gauss' law this may be re-written as
\[-\int_{S} P(\xi,t)\hat{n}dS = -\int_{V}\partial_{\xi}P(\xi,t)dV = -\int_{a}^{x}\partial_{\xi}P(\xi,t)A(\xi,t)d\xi.\]
This accounts for all the forces acting on the blood contained within the artery between axial locations $a$ and $x.$

Having established the rate of change of momentum, along with the various contributing forces, we  require an equation of motion for this finite fluid element. On applying Newton's second law to the above acceleration and forces, we obtain,
\begin{align*}
    &\tilde{\rho}\partial_{t}\int_{a}^{x}u(\xi,t)A(\xi,t)d\xi\\ =& \tilde{\rho}\int_{a}^{x}[u(\xi,t)u_{\xi}(\xi,t)A(\xi,t) - u(\xi,t)(u(\xi,t)A(\xi,t))_{\xi}]d\xi - \int_{a}^{x}\partial_{\xi}P(\xi,t)A(\xi,t)d\xi.
\end{align*}
Dividing both sides by $\tilde{\rho},$ operating on the resulting equation with $\partial_{x},$ and applying the fundamental theorem of calculus to both sides we obtain,
\begin{align*}
    &(u(x,t)A(x,t))_{t}\\=& u(x,t)u_{x}(x,t)A(x,t) - u(x,t)(u(x,t)A(x,t))_{x} - \frac{1}{\tilde{\rho}}P_{x}(x,t)A(x,t)
\end{align*}
Expanding the derivatives on each side, applying the continuity equation (\ref{gov.1}), and dividing by $A(x,t),$ we find
\begin{equation}\label{Euler}
    u_{t}(x,t) + u(x,t)u_{x}(x,t) + \frac{1}{\tilde{\rho}}P_{x}(x,t) = 0.
\end{equation}
We see that in our model the blood flow will satisfy \textit{Euler's Equation} for an incompressible, inviscid and irrotational fluid.

In general it is difficult to obtain from first principles an explicit expression for the transmural pressure $P(x,t).$ However a large body of experimental data exists to suggest a plausible correspondence between the pressure and the cross section of the tube itself. Indeed, a standard example is the so called \textit{Windkessel} model \cite{KeenerSneyd2009}, in which the pressure is related linearly to the cross-section,
\[P(x,t) = P(A(x,t)) = k_WA(x,t).\]
The constant $k_{W}$ relates the elastic restoring force of the tube when distended to cross-section $A(x,t),$ to the pressure exerted on the blood. This is generally determined from clinical data.
In this paper we will adopt a slightly more complicated model, whereby the relationship between pressure and aortic cross-section is quadratic, namely
\begin{equation*}
    P(x,t) = P(A(x,t)) = kA^{2}(x,t).
\end{equation*}
Again, the constant $k$ in our  model is determined from clinical data, however it shall play no significant role in any of the remaining results. As such, we shall redefine our cross-section, such that $\frac{k}{\tilde{\rho}}A^{2}(x,t)\to \frac{1}{2}A^{2}(x,t).$ Finally the second governing equation of our model for arterial blood flow is,
\begin{equation}\label{gov.2}
    u_{t}(x,t) + u(x,t)u_{x}(x,t) + A(x,t)A_{x}(x,t) = 0.
\end{equation}
The equations (\ref{gov.1}) and (\ref{gov.2}) when taken together constitute a two-component dispersionless Burgers equation.

\section{Solutions when $k > 0$}
In this section we consider in more detail the behaviour of the system,
\begin{eqnarray}\label{Burgers.2}
      \nonumber u_t + uu_x + AA_x &=& 0\\
                A_t + (uA)_x &=& 0,
\end{eqnarray}
where we have chosen a definite sign for the physical parameter $k = 1.$ As we already mentioned, one of the great advantages of this system is that it may be solved directly by the method of characteristics. To illustrate this we shall define a pair of diffeomorphisms $\diff(x,t)$ by the following criteria:
\begin{eqnarray}\label{diff.1}
    \nonumber\partial_t\psi_{\pm}(x,t) &=& u(\psi_{\pm},t) \pm A(\diff,t),\\
    \diff(x,0) &=& x.
\end{eqnarray}
When we apply $\partial_{t}$  once more to (\ref{diff.1}), and using (\ref{Burgers.2}) we find,
\begin{equation}\label{second_derivative}
\frac{\partial^2\diff}{\partial t^2}(x,t) = 0.
\end{equation}
Since $\partial^{2}_{t}\psi_{\pm}(x,t) = 0,$ it follows that $\partial_{t}\psi_{\pm}(x,t)$ must be constant in time, and may depend on $x$ only.

Using this, and the definition supplied in (\ref{diff.1}), we see that the physical variables evaluated along the flow of $\psi_{\pm}(x,t)$ must satisfy the constraints,
\begin{equation}\label{diff.3}
    u(\diff, t) \pm A(\diff, t) = u_0(x) \pm A_{0}(x).
\end{equation}
We see that (\ref{second_derivative}) requires $\diff(x,t)$ be linear in $t,$ while the second equation in (\ref{diff.1}) imply that $\diff(x,t)$ satisfy,
\begin{equation*}
    \diff(x,t) = x + t\gamma_{0}^{(\pm)}(x).
\end{equation*}
Applying $\partial_{t}$ to $\diff(x,t)$ as it is given in this expression, and comparing to (\ref{diff.1}) and (\ref{diff.3}), we can see that the functions $\gamma_{0}^{(\pm)}(x)$ may be written in terms of the initial data,
\[\gamma_{0}^{(\pm)}(x) = u_{0}(x)\pm A_{0}(x).\]
So we see that we may solve (\ref{diff.1}) to find,
\begin{equation}\label{diff.2}
    \psi_{\pm}(x,t) = x + t(u_{0}(x)\pm A_{0}(x)).
\end{equation}
allowing us to express $\diff(x,t)$ in terms of the given initial data $u_{0}(x)$ and $A_{0}(x).$

Since  are diffeomorphisms, at least for appropriately chosen initial data, it is in principle possible to invert (\ref{diff.2}), and upon doing so one may obtain explicit solutions for the physical variables $u(x,t)$ and $A(x,t)$ in terms of the initial data $(u_{0}(\cdot), A_{0}(\cdot)):\mathbb{R}^{2}\to\mathbb{R}^{2},$
\begin{eqnarray}\label{u_and_A}
    \nonumber u(x,t) &=& \frac{1}{2}\left[u_{0}(\psi_{+}^{-1}) + u_{0}(\psi^{-1}_{-})\right] + \frac{1}{2}\left[A_{0}(\psi_{+}^{-1}) - A_{0}(\psi^{-1}_{-}) \right],\\
    A(x,t) &=& \frac{1}{2}\left[u_{0}(\psi_{+}^{-1}) - u_{0}(\psi^{-1}_{-}) \right] + \frac{1}{2}\left[A_{0}(\psi_{+}^{-1}) + A_{0}(\psi^{-1}_{-}) \right]
\end{eqnarray}
where it is understood that $\psi^{-1}_{\pm} := \psi^{-1}_{\pm}(x,t).$

\smallskip
\noindent
{\underline{Example:}} As an example, we will consider a solution in which our initial data is of the form $u_{0}(x)\sim A_{0}(x) \sim x^{\frac{1}{3}}.$ Specifically,  our initial data is defined by,
\begin{equation}\label{ID.1}
    u_{0}(x) \pm A_{0}(x) = a_{\pm}x^{\frac{1}{3},}
\end{equation}
where $a_{\pm}$ are constants. It follows from (\ref{diff.2}) that the diffeomorphisms $\diff(x,t)$ may be written as,
\begin{equation}\label{diff-ex.1}
    \psi_{\pm}(x,t) = x + a_{\pm}tx^{\frac{1}{3}}.
\end{equation}
Making the invertible substitution $x\to w^{3},$  and with the corresponding change of variables  $\psi_{\pm}(x,t)\rightarrow y_{\pm}(w,t),$ we may re-write the diffeomorphisms in (\ref{diff-ex.1}),
\begin{equation*}
    y_{\pm}(w,t) = w^{3} + a_{\pm}tw.
\end{equation*}
We now have a pair of monic polynomials with argument $w\in\R,$ namely,
\begin{equation}\label{polynomial}
    w^{3} + a_{\pm}tw - y_{\pm}^{3}(w,t) = 0,
\end{equation}
the discriminants of which are given by,
\begin{equation}\label{discriminant}
    D = -4(a_{\pm}t)^{3} - 27.
\end{equation}
The values of the discriminant determine the quantity and nature of solutions for $w.$ In particular, we are interested in real solutions $w(y_{\pm},t).$

Equivalently, this solution allows us to solve for $x$ in terms of $\psi_{\pm}(x,t)$ and $t,$ that is, it offers us an expression for $\psi_{\pm}^{-1}(x,t).$
Depending on the discriminant $D,$ we may have several real roos, all of which may be distinct, or some of which may be equal. In the case $D=0,$  all roots are equal and real, for $a_{\pm} \in \mathbb{R}$ and real $\psi_{\pm}(x,t).$  Moreover, the discriminant allows us to explicitly calculate the time at which the wave breaking occurs,
\begin{equation}
    T_{\pm} =-\frac{3}{\sqrt[3]{4}a_{\pm}}.
\end{equation}
In this case, the functions $\psi_{\pm}(x,t)$ do not have unique inverses,  and so no longer behave as diffeomorphisms. Furthermore, the corresponding solutions $u(x,t)$ and $A(x,t),$as defined by (\ref{u_and_A}), will no longer satisfy $|u_{x}(x,t)| < \infty$ and $|A_{x}(x,t)| < \infty,$ since $\psi_{\pm}(x,t)$ are no longer strictly monotone increasing functions of $x,$ for all $t > 0.$

\textbf{\textit{Remark}}
The functions $\psi_{\pm}(x,t)$ are diffeomorphisms only in the case where wave-breaking does not occur. We assume $(u_{0}(x),A_{0}(x))\in C^{1}\times C ^{1}$ and also, $u_{0}(x)$ and $A_{0}(x)$ are bounded, that is,
\[\displaystyle{\sup_{x\in\mathbb{R}}}(|u_{0}(x)| + |A_{0}(x)|) < \infty.\]
Therefore, our solutions blow-up only if $u_{0}'(x) \pm A_{0}'(x) < 0$ at some point and $t > 0,$ otherwise the solutions are global.

\section{Solutions when $k < 0$}
A related system is the so called wrong sign Burgers equation given by,
\begin{eqnarray}\label{Burgers2.1}
    \nonumber   u_t + uu_x - (A^2)_x &=& 0\\
                A_t + (uA)_x &=& 0,
\end{eqnarray}
which differs from (\ref{b2_1}, \ref{b2_2}) by the sign of the $(A^{2})_{x}$ term. In this case we set $k = -1.$
It can be shown that the Hamiltonian for the system may be written as,
\begin{equation}\label{Burgers.2H}
    H = \frac{1}{2}\int_{-\infty}^{\infty}(u^{2}(x,t) - A^{2}(x,t))dx,
\end{equation}
which we can see does not have a definite sign. It follows that the system appearing in (\ref{Burgers2.1}) will describe physically unstable systems.
In analogy to the case of solution via characteristics, we may construct a pair complex conjugate mappings $\chi_{\pm}(x,t),$ which are formally defined by,
\begin{equation}
    \partial_{t}\chi_{\pm}(x,t) = u(\chi_{\pm},t) \pm iA(\chi_{\pm},t).
\end{equation}
As in the case of (\ref{diff.3}), we find the analogous relations,
\begin{equation}\label{diffRI}
    u(\chi_{\pm},t) \pm iA(\chi_{\pm},t) = u_{0}(x) \pm iA_{0}(x).
\end{equation}
The construction of such a pair of solutions was given in \cite{Kodama02}, wherein the authors obtained solutions by the method of Riemann invariants, with the mappings $\chi_{\pm}(x,t)$ given by
\begin{eqnarray}\label{Riemann1}
    \chi_{\pm}(x,t) &=& x - t(1 \pm i\sqrt{3})\left(\frac{x}{\lambda}\right)^{\frac{1}{3}}\\
    &=& x - 2te^{\pm\frac{i\pi}{3}}\left(\frac{x}{\lambda}\right)^{\frac{1}{3}},
\end{eqnarray}
It follows that the initial data corresponding to each of these mappings is given by,
\begin{equation}\label{Riemann2}
    u_{0}(x) = -\left(\frac{x}{\lambda}\right)^{\frac{1}{3}} \quad   A_{0}(x) = \mp \left(\frac{x}{\lambda}\right)^{\frac{1}{3}},
\end{equation}
with the parameter $\lambda$ being freely adjustable.

The corresponding solutions are found to be,
\begin{equation}\label{Riemann3}
    A(\chi_{\pm}(x,t),t) = \sqrt{\frac{8t}{\lambda} + 3u^{2}(\chi_{\pm}(x,t),t)},
\end{equation}
where the solution $u(\cdot,t)$ is the real root of the cubic polynomial,
\begin{equation}\label{Riemann4}
    \lambda u^{3}(\xi,t) + 2tu(\xi,t) + \xi = 0,
\end{equation}
for arbitrary $\xi\in\mathbb{R}.$

Having an explicit expression for $A(\chi_{\pm},t)$ in term of $u(\chi_{\pm},t)$ we now require an explicit expression for $u(\chi_{\pm},t)$ in terms of $x$ and $t.$ To proceed, we notice from (\ref{diffRI}) that our solution $u(\chi_{\pm}(x,t),t)$ must satisfy
\begin{equation}\label{Riemann5}
    u(\chi_{\pm}(x,t)) = u_{0}(x) \pm iA_{0}(x) \mp iA(\chi_{\pm}(x,t),t).
\end{equation}
It follows from (\ref{Riemann2}) and (\ref{Riemann3}) that the solutions $u(\chi_{\pm}(x,t),t)$ may be written as,
\begin{equation}
    u(\chi_{\pm}(x,t),t) = -2e^{\pm\frac{i\pi}{3}}\left(\frac{x}{\lambda}\right)^{\frac{1}{3}} \mp\sqrt{\frac{8t}{\lambda} + 3u^{2}(\chi_{\pm}(x,t),t)}.
\end{equation}
Substituting this expression into the cubic polynomial in (\ref{Riemann4}), we find,
\begin{equation}\label{Riemann6}
    u(\chi_{\pm}(x,t),t) = -\frac{e^{\pm i\frac{2\pi}{3}}}{2}\left(\frac{x}{\lambda}\right)^{\frac{1}{3}} \pm \sqrt{-\frac{3e^{\pm i\frac{\pi}{3}}}{8}\left(\frac{x}{\lambda}\right)^{\frac{2}{3}} - \frac{t}{\lambda}},
\end{equation}
and so we also have explicit solutions for $A(\chi_{\pm},t)$ in terms of $x$ and $t,$
as follows from (\ref{Riemann3}).

\section{Solutions that are not continuous diffeomorphisms}
The phenomenon of wave-breaking was one of the most interesting and exciting aspects of the Camassa-Holm equation to be investigated following its discovery in the work \cite{CH93}. In this article we have focussed on a similar though somewhat simplified system to model the flow of blood in arteries, and we would like to investigate the possibility of such a phenomenon arising in the context of this model. The clinical relevance of this is in relation to the phenomenon of the \textit{pistol shot pulse.} Such behaviour arises during systole of patients with \textit{aortic insufficiency}, when blood ejected into the artery is regurgitated back through the aortic valve, into the ventricles. To maintain systole pressure, the heart will contract more during ventricular systole, thereby ejecting blood into the artery with greater pressure. After systole, the ejected blood will induce a pressure gradient along the radial artery, however, owing to the initial aortic regurgitation, there will be insufficient blood mass to maintain the pressure throughout the length of the artery. This in turn will cause a sudden increase followed by a rapid decrease in the aortic cross section, which is observed as the femoral pistol-shot pulse.

In this section we aim to establish the conditions under which a sudden expansion in the aortic cross section is followed by a collapse thereof, in rapid succession. We aim to show that the phenomena of wave-breaking arising in the system (\ref{b2_1})-(\ref{b2_2}), which we are using as a simplified model of aortic blood flow, is sufficient to account for this clinical phenomenon. Mathematically speaking, wave breaking occurs when our solutions $u(x,t)$ and $A(x,t)$ remain bounded for all $(x,t) \in \mathbb{R}\times[0,T),$ while the magnitude of their gradients become singular in finite time \cite{CE98, Whitham},
\[|u_{x}(x,t)| + |A_{x}(x,t)| \to \infty \quad t\to T.\]
We begin by constraining our initial data, such that,
\[(u_{0}(x),A_{0}(x))\in H^{1}\times H^{1} \subset C^{1}\times C^{1}.\]
We will show that continuous solutions $u(x,t)$ and $A(x,t)$ do indeed remain finite, if their slopes remain finite. To do so, we first multiply each member of the system in (\ref{Burgers.2}) by $u(x,t)$ and $A(x,t)$ respectively. Then integrating over $\mathbb{R}$ and imposing the boundary conditions,
\begin{equation}\label{wb.1}
\displaystyle\lim_{x\to\pm\infty}u(x,t) = 0, \quad \displaystyle\lim_{x\to\pm\infty}A(x,t) = 0,
\end{equation}
we find the following conditions
\begin{eqnarray}\label{wb.2}
    \nonumber\frac{1}{2}\frac{d}{dt}\int u^2dx &=& \frac{1}{2}\int u_xA^2dx\\
    \frac{1}{2}\frac{d}{dt}\int A^2dx &=& -\frac{1}{2}\int u_xA^2dx.
\end{eqnarray}
Indeed we see from this result that the quantity,
\[H = \int(u^{2}(x,t) + A^2(x,t))dx,\]
is an integral of motion, and may act as a Hamiltonian for our system.

Next we operate with $\partial_x$ on each member of (2), and multiply by $u_x(x,t)$ and $A_x(x,t)$ respectively. Imposing the boundary conditions (\ref{wb.1}) and integrating over $\mathbb{R},$ we find,
\begin{eqnarray}\label{wb.3}
    \nonumber\frac{1}{2}\frac{d}{dt}\int u_{x}^{2}dx &=& -\frac{1}{2}\int u_{x}^{3}dx - \int u_{x}A_{x}^{2}dx - \int u_{x}AA_{xx}dx\\
    \frac{1}{2}\frac{d}{dt}\int A_{x}^{2}dx &=& \int u_xAA_{xx}dx - \frac{1}{2}\int u_{x}A_{x}^{2}dx.
\end{eqnarray}
Using (\ref{wb.2}) and (\ref{wb.3}), we find that,
\begin{equation}\label{wb.4}
    \frac{1}{2}\frac{d}{dt}\int(u^2 + u_{x}^{2} + A^{2} + A_{x}^{2})dx = -\frac{1}{2}\int u_{x}(u_{x}^{2} + 3A_{x}^{2})dx.
\end{equation}
It follows from (\ref{wb.4}) that
\begin{equation}\label{wb.5}
    \frac{d}{dt}\int(u^{2} + u_{x}^{2} + A^{2} + A_{x}^{2})dx \leq 3M_{1}\int(u^{2} + u_{x}^2 + A^{2} + A_{x}^{2})dx,
\end{equation}
where $\displaystyle\sup_{x\in\mathbb{R}}|u_{x}(x,t)| = M_1.$

Gronwall's inequality \cite{Bauer_Nohel} states that for a pair of   functions $f(t)$ and $g(t),$ continuous on some interval $\alpha\leq t < \beta,$ and with $f(t)$ differentiable on $(\alpha,\beta),$  and such that
\begin{eqnarray*}
    f'(t) &\leq& f(t)g(t), \quad t \in (\alpha, \beta),\\
    &&\mbox{ then }\\
    f(t) &\leq& f(\alpha)\exp\left(\displaystyle\int_{\alpha}^{t}g(\xi)d\xi\right).
\end{eqnarray*}
Upon applying Gronwall's inequality to (\ref{wb.5}), we find that
\begin{equation}\label{wb.6}
    \int(u^{2} + u_{x}^{2} + A^{2} + A_{x}^{2})dx \leq K_1e^{3M_{1}t} < \infty:  \quad t \in [0,T)
\end{equation}
Here we introduce $K_1 = \displaystyle\int_{\mathbb{R}}(u_{0}^2 + u_{0}'^{2} + A_{0}^{2} + A_{0}'^{2})dx,$ with $u_{0}(x) = u(x,0)$ and $A_{0}(x) = A(x,0).$
In addition, for any function $u(x,t) \in H^{1},$ we have
\begin{equation*}
    |u(x,t)|^{2} \leq \|u\|_{1} = \displaystyle\int_{\mathbb{R}}(u^{2}(\xi,t) + u_{\xi}^{2}(\xi,t))d\xi.
\end{equation*}
Since we assumed $(u(x,t),A(x,t)) \in H^{1}(\mathbb{R})\times H^{1}(\mathbb{R}),$ it follows from this inequality, along with the result in (\ref{wb.6}) that
\begin{equation}\label{wb.7}
    |u(x,t)|^{2} + |A(x,t)|^{2} \leq \|u\|_{1} + \|A\|_{1} \leq K_{1}e^{3M_{1}t}, \quad (x,t)\in\mathbb{R}\times[0,T),
\end{equation}
It follows that the solutions $|u(x,t)|$ and $|A(x,t)|$ remain bounded for all $(x,t)\in\mathbb{R}\times[0,T)$ if $|u_{x}(x,t)|$ and $|A_{x}(x,t)|$ remain bounded for the same values of $x$ and $t.$

Next, we would like to establish under what conditions the functions $u_{x}(x,t)$ and $A_{x}(x,t)$ actually become singular, while $u(x,t)$ and $A(x,t)$ remain bounded. We return to the diffeomorphisms introduced in Section 3, in particular (\ref{diff.3}). Differentiating this once with respect to $x,$ we find
\begin{equation}\label{blow-up.1}
    [u_{x}(\psi_{\pm},t) \pm A_{x}(\psi_{\pm},t)]\cdot\psi_{\pm,x}(x,t) = u_{0}'(x) \pm A_{0}'(x),
\end{equation}
Substituting (\ref{diff.2}) into this, we find
\begin{equation}\label{blow-up.2}
    u_{x}(\psi_{\pm},t) \pm A_{x}(\psi_{\pm},t) = \frac{u_{0}'(x) \pm A_{0}'(x)}{1 + t\cdot(u_{0}'(x) \pm A_{0}'(x))}.
\end{equation}
Under the condition $(u_{0}(x),A_{0}(x))\in H^{1}\times H^{1},$
and with $\displaystyle{\inf_{x\in\mathbb{R}}}[u_{0}'(x)\pm A_{0}'(x)] < 0,$ it follows that
\begin{equation}
    u_{x}(\psi_{\pm},t)\pm A_{x}(\psi_{\pm},t) \to -\infty, \quad t \to \displaystyle{\inf_{x\in \mathbb{R}}}\frac{1}{|u_{0}'(x)\pm A_{0}'(x)|}.
\end{equation}
It follows that with these initial conditions the system (\ref{b2_1})-(\ref{b2_2}) develops wave-breaking phenomena.

Conversely, for $u_{0}(x) \pm A_{0}(x) > 0, \ x \in \R,$ we see that the denominator is non-zero for all $t > 0.$ In this case the functions $\psi_{\pm}(x,t)$ are invertible for all $(x,t)\in\R\times[0,\infty).$ In this case the functions in (\ref{blow-up.2}) remain bounded, so that
\begin{equation}
    |u_{x}(\psi_{\pm},t) \pm A_{x}(\psi_{\pm},t)| < \infty, \ t > 0
\end{equation}
So the solutions $u(\cdot,t)$ and $A(\cdot,t)$ are global, if our initial data  satisfies $u_{0}'(x) \pm A_{0}'(x) > 0, \ x \in \R.$

\section{Conclusion}
We have derived a model for the flow of blood through an artery. We model the blood as a Newtonian fluid, flowing through an elastic tube. The restoring forces in the elastic tube are assumed to depend on the distension or contraction of the artery from its equilibrium cross section. Specifically, the pressure exerted on the blood by the artery is proportional to the square of the cross-section.

One particular nonlinear effect we investigate in the model is the steepening of the wave front as it moves away from the ventricle during systole. The same phenomenon was studied in other models by several authors, in particular in the works \cite{Aniker_a, Aniker_b, Lighthill_1}.

In the event that the wave front becomes too steep, the top of the wave front overtakes the bottom, and a shock (or discontinuity) develops as a solution, which is typical of hyperbolic equations.
Of course a true shock is not possible in this context as blood viscosity and the elastic properties of the arterial wall preclude the formation of a discontinuous solution. Nevertheless, it is possible to generate very steep pressure gradients within the artery, due to aortic regurgitation during systole. Under such circumstances, clinicians report the phenomenon of the pistol shot pulse, in which the arterial wall undergoes an abnormal expansion, followed by a sudden collapse, which we interpreted here as the development of shock in the wave-front\cite{KeenerSneyd2009}.

\medskip
Received May 2011; revised June 2011.


\begin{thebibliography}{99}

\bibitem{Ach1990}
D. J. Acheson,
``Elementary Fluid Dynamics,"
Oxford University Press, Oxford, 1990.

\bibitem{Aniker_a}
M. Aniker, R. L. Rockwell and E. Ogden,
\emph{Nonlinear analysis of flow pulses and shock waves in arteries. Part I: derivation and properties of mathematical model},
Zeitschrift f\"{u}r Angewandte Mathematik und Physik., \textbf{22} (1971a), 217--246.

\bibitem{Aniker_b}
M. Aniker, R. L. Rockwell and E. Ogden,
\emph{Nonlinear analysis of flow pulses and shock waves in arteries. Part II: parametric study related to clinical problems},
Zeitschrift f\"{u}r Angewandte Mathematik und Physik., \textbf{22} (1971b), 563--581.

\bibitem{Bauer_Nohel}
F. Bauer and J. A. Nohel,
``The Qualitative Theory of Ordinary Differential Equations: An Introduction,"
W.A Benjamin, New York, 1969.

\bibitem{Beals_et_al}
R. Beals, D. Sattinger and J. Szmigielski,
\emph{Accoustic scattering and the extended KdV hierachy},
Adv. Math., \textbf{140} (1998), 190--206 {doi: 10.1005/aima.1998.1768}.

\bibitem{BKST09}
A Boutet de Monvel, A. Kostenko, D. Shepelsky and G. Teschi,
\emph{Long-time asymptotics for teh Camassa-Holm equation},
SIAM J. Math. Anal., \textbf{41} (2009), 1559--1588 {doi: 10.1137/090748500}.

\bibitem{CH93}
R. Camassa and D. D. Holm,
\emph{An integrable shallow water equation with peaked solitons},
Phys. Rev. Lett., \textbf{71} (1993), 1661--1664 {doi: 10.1103/PhysRevLett.71.1661}.

\bibitem{CLZ05}
M. Chen, S.-Q. Liu and Y. Zhang,
\emph{A two-component generalization of the Camassa-Holm equation and its solutions},
Lett. Math. Phys., \textbf{75} (2006), 1--15; nlin.SI/0501028.

\bibitem{CL10}
R. M. Chen and Y. Liu,
\emph{Wave breaking and global existence for a generalized two-component Camassa-Holm system},
International Mathematics Research Notices, Article ID rnq118 (2010), 36 pages; doi10.1093/imrn/rnq118.

\bibitem{CE98}
A. Constantin and J. Escher,
\emph{Wave breaking for nonlinear nonlocal shalow water equation},
Acta. Math., \textbf{181} (1998), 229--243 {doi: 10.1007/BF02392586}.

\bibitem{CGI06}
A. Constantin, V. Gerdjikov and R. I. Ivanov,
 \emph{Inverse scattering transform for the Camassa -Holm equation},
 Inverse Problems, \textbf{22} (2006), 2197--2207 {doi: 10.1088/0266-5611/22/6/017}.

\bibitem{CI08}
 A. Constantin and R. Ivanov,
 \emph{On an integrable two-component Camassa-Holm shallow water system},
Phys. Lett. A, \textbf{372} (2008),  7129--7132 {doi: 10.1016/j.physleta.2008.10.050}.

\bibitem{CJ2008}
A. Constantin and R. S. Johnson,
\emph{Propagation of very long water waves, with vorticity, over variable depth, with applications to tsunamis},
Fluid Dynam Res., \textbf{40} (2008), 175--211.

\bibitem{CL09}
A. Constantin and D. Lannes,
\emph{The hydro-dynamical relevance of teh Camassa-Holm and Degasperis-Proceisi equations},
Arch. Ration. Mech. Anal., \textbf{192} (2009), 165--186.

\bibitem{CMK99}
A. Constantin and H. P. McKean,
\emph{A shallow water equation on the circle},
Commun. Pure Appl. Math., \textbf{52} (1999), 949--982.

\bibitem{CS00}
A. Constantin and W. Strauss,
\emph{Stability of a class of solitary waves in compressible elastic rods},
Phys. Lett. A, \textbf{270} (200), 140--148.

\bibitem{Dai98}
H. H. Dai,
\emph{Model equations for nonlinear dispersive waves in a compressible Mooney-Rivlin rod},
Acta. Math., \textbf{127} (1998), 193--207.

\bibitem{Escher07}
J. Escher, O. Lechtenfeld and Z. Yin,
\emph{Well-posedness and blow-up phenomena for the 2-component Camassa-Holm equation},
Discrete Contin. Dyn. Syst., \textbf{19} (2007), 493--513.

\bibitem{Escher08}
   J. Escher and Z. Yin,
   \emph{Well-posedness, blow-up phenomena and global solutions for the \emph{b}-equation},
    J. Reine Angew. Math., \textbf{624} (2008), 51--80 {doi: 10.1515/CRELLE.2008.080}.

\bibitem{F06}
G. Falqui,
\emph{On a Camassa-Holm type equation with two dependent variables},
J. Phys. A, \textbf{39} (2006), 327--342 {doi: 10.1088/0305-4470/39/2/004}.

\bibitem{GL10}
G. Gui and Y. Liu,
\emph{On the global existence and wave-breaking criteria for the two-component Camassa-Holm system},
J. Funct. Anal., \textbf{258} (2010), 4251--4278 {doi: 10.1016/j.jfa.2010.02.008}.

\bibitem{H09}
D. Henry,
\emph{Infinite propagation speed for a two component Camassa-Holm equation},
Discrete Contin. Dyn. Syst. Ser. B, \textbf{12} (2009) 597--606 {doi: 10.3934/dcdsb.2009.12.597}.

\bibitem{HT09}
D. D. Holm and C. Tronci,
\emph{Geodesic Vlasov equations and their integrable moment closures},
J. Geom. Mech., \textbf{1}  (2009), 181--208.

\bibitem{HI11}
D. D. Holm and R. I. Ivanov,
\emph{Two-component CH system: inverse scattering, peakons and geometry},
Inverse Problems, \textbf{27} (2011), 045013; doi: 10.1088/0266-5611/27/4/045013.

\bibitem{I06}
R. I. Ivanov,
\emph{Extended Camassa-Holm hierarchy and conserved quantities},
Z. Naturforsch., \textbf{61a} (2006),  133--138; nlin.SI/0601066.

\bibitem{I09}
R. I. Ivanov,
\emph{Two-component integrable systems modelling shallow water waves: the constant vorticity case},
 Wave Motion, \textbf{46}  (2009), 389--396.

\bibitem{Johnson02}
R. S. Johnson,
\emph{Camassa-Holm, Kortweg-de Vries and related models for water waves},
J. Fluid Mech., \textbf{455} (2002), 63--82 {doi: 10.1017/S0022112001007224}.

\bibitem{KeenerSneyd2009}
J. Keener and J. Sneyd,
``Mathematical Physiology 1: Cellular Physiology,"
Springer, 2009.

\bibitem{Kodama02}
Y. Kodama and B. Konopelchenko,
\emph{Singular sector of the Burgers-Hopf hierarchy and deformations of hyperelliptic curves},
J. Phys. A: Math. Gen., \textbf{35} (2002), L489--L500.

\bibitem{Lighthill_1}
J. Lighthill,
``Mathematical Biofluiddynamics," SIAM, Philadelphia, PA. 1975.

\bibitem{LZ05}
S.-Q. Liu and Y. Zhang,
\emph{Deformations of semisimple bi-Hamiltonian structures of hydrodynamic type},
J. Geom. Phys., \textbf{54} (2005), 427--53 {10.1088/0305-4470/35/31/104}.

\bibitem{SA}
P. Olver and P. Rosenau,
\emph{Tri-Hamiltonian duality between solitons and solitary-wave solutions having compact support},
Phys. Rev. E, \textbf{53}  (1996), 1900--1906 {10.1103/PhysRevE.53.1900}.

\bibitem{Pedley}
T. J. Pedley,
\emph{Blood flow in arteries and veins},
 in ``Perspectives in Fluid Dynamics," Cambridge University Press, 2003.

\bibitem{Strauss1992}
W. A. Strauss,
``Partial Differential Equations: An Introduction,"
John Wiley \& Sons Inc., 1990.

\bibitem{Whitham}
G. B Whitham,
``Linear and Nonlinear Waves,"
John Wiley \& Sons, New York, 1980.

\end{thebibliography}
\end{document}